\documentclass[fleqn,twoside]{article}
\usepackage{espcrc2}

\usepackage{graphicx}
\usepackage[figuresright]{rotating}

\newcommand{\AmS}{{\protect\the\textfont2
  A\kern-.1667em\lower.5ex\hbox{M}\kern-.125emS}}

\def\text{~}

\def\hc2{(\hbar c)^2}

\def\r2{\langle r^2 \rangle}
\def\Q2{$Q^2$}

\def\gev2c2{GeV$^2$/$c^2$}
\def\fm2{\text{fm}^2}

\def\a2         {{\mbox{$a_{2}                                        \
$}}}

\def\a2pigamma  {{\mbox{$a_2^-\rightarrow \pi^-\gamma                \
$}}}

\title{Color Transparency via Coherent Exclusive $\rho^0$
Production}

\author{M. Moinester\address[TAU]{School of Physics and Astronomy, 
R. and B. Sackler Faculty of Exact Sciences, Tel Aviv University, 69978 Ramat Aviv,
Israel},
O. A. Grajek\address[SI]{So\l tan Institute for Nuclear Studies, 
ul. Ho\.{z}a 69, PL 00-681 Warsaw, Poland},
E. Piasetzky\addressmark[TAU],
A. Sandacz\addressmark[SI]
\thanks{The authors gratefully acknowledge useful
discussions with L. Frankfurt and M. Strikman.
The work was supported in part by 
the Israel Science Foundation (ISF) founded by the Israel Academy of
Sciences and Humanities, Jerusalem, Israel and
by the Polish State Committee for
Scientific Research (KBN SPUB Nr 621/E-78/SPUB-M/CERN/P-03/DZ 298/2000
and KBN grant Nr 2 P03B 113 19). One of us (A.S.) acknowledges  
support from the Raymond and Beverly Sackler Visiting Chair in Exact Sciences
during his stay at Tel Aviv 
University.}
}
\begin{document}
\begin{abstract}
 We examine the potential of the COMPASS experiment at CERN to study color transparency
via exclusive coherent vector meson production in hard muon-nucleus
scattering.
It is demonstrated that COMPASS has high sensitivity to test this
important prediction of perturbative QCD.
\vspace{1pc}
\end{abstract}

\maketitle

\section{INTRODUCTION}

One prediction of pQCD is that high $Q^2$
longitudinally polarized virtual photons $\gamma^{\ast}_{L}\/$
fluctuate into hadronic components, e.g. $q\bar{q}$ pairs,
whose transverse size 
$b = \linebreak \: \mid \! \bar{r}_{\! \perp q} - \bar{r}_{\! \perp \bar{q}} \! \mid $ 
decreases with $Q^2 \!$, \,$b^2 \propto (1/Q^{2}) \!$.
\,At large $Q^2$ the values of $b$ are 
significantly smaller than the size of the nucleon.
Such a Small Size Configuration (SSC)
interacts with hadrons with small cross sections, a phenomenon known as  
{\it Color Transparency\/} (CT) \cite {bfgms,fms,frs}.

Cross section for the interaction of SSC with a hadron target has
been calculated in QCD using a factorization theorem
\cite{fms,frs}. Observing CT in particular kinematics would
prove experimentally the applicability of the QCD factorization
theorem for those kinematics. Such a proof provides important
complementary support to a class of spin physics experiments, for
example
measurements of generalized parton distributions
\cite {jp}.

The prerequisite for observing CT is to select a sample containing SSC
mini-mesons
via large $Q^{2}\!$, \,high $p^{}_{t}\/$,
or large produced mass.      
For hard exclusive $\rho ^0$ leptoproduction,
in addition to large $Q^2$, selection of the longitudinally
polarized mesons is required.
In order to clearly observe CT, it is also necessary that the SSC 
lives long enough  
while propagating through the nucleus.
This requirement is characterized in terms of 
the coherence length $l_c$ \cite {RPW,HERMES}.

Strong recent evidence for CT comes from
Fermilab E791 experiment on the \mbox{$A$-dependence}
of coherent diffractive dissociation of pions into two high-$p^{}_{t}\/$ jets
\cite{Wei97}. Also the E691 
\cite{Sokolov} 
and NMC \cite {NMCJpsi} results on $A$-dependence of coherent
$J \! / \! \psi\/$ photon and muon production 
are consistent with CT. 
The HERMES \cite {HERMES}, NMC \cite {NMC}, and E665 \cite
{E665} leptoproduction experiments 
have made extensive studies of  CT via incoherent 
vector meson production 
(VMP), most of the data for $\rho^0$ production. 
The 
BNL (p,2p) 
\cite {ac}
and SLAC (e,e'p)  
\cite {NE18} 
high four momentum transfer 
quasi-elastic data  
were also compared to CT predictions. 

We proposed \cite {asct}  to study CT at COMPASS 
\cite {compass,bradamante} via {\it exclusive
coherent VMP\/} via reactions such as  
$\mu A \rightarrow \mu \, A \rho$  
on C and Pb nuclei. 
The selections of coherent or incoherent production will
be mainly using the $t$-distribution;
at the lowest $\mid \! t \! \mid $ values coherent events predominate.

The basic observable for each process studied
is the ratio of the nuclear transparencies
for lead and carbon, $R^{}_{\rm T} = T_{\rm{Pb}}/T_{\rm{C}} =
(\sigma _{\rm{Pb}}/A_{\rm{Pb}})/(\sigma _{\rm{C}}/A_{\rm{C}})$. 
T is the ratio of the cross section per nucleon 
on a nucleus $A$ to the corresponding cross section on a free nucleon.

\section{SIMULATION OF EXCLUSIVE $\rho^{0}$ EVENTS}
Simulations were carried out  \cite {asct} with a dedicated fast Monte Carlo
program which generates deep inelastic exclusive coherent $\rho^{0}$
events with subsequent decay $\rho ^0 \rightarrow \pi^{+} \pi^{-} \!$.
Experimental effects such as  
acceptance, kinematic smearing, etc., were included.
The differential cross sections for the proton target
$\frac{{\rm{d}} \sigma^{}_{\! N}}{{\rm{d}} t}$
is related to that for coherent production on the nucleus $A$ by:

\begin{equation}  \mbox{\Huge (} \frac{{\rm{d}} \sigma^{\rm
coh}_{\! \! A}}{{\rm{d}} t}   \mbox{\Huge )}_{\! \! i} =   A^{2}_{\rm
eff \!, \; coh} \cdot e^{< R^{2}_{\! A} > \, t/3} \cdot   \mbox{\Huge
(} \frac{{\rm{d}} \sigma^{}_{\! N}}{{\rm{d}} t} \mbox{\Huge
)}^{}_{\! \! i} \;\: 
\label{EQ_MC50}. 
\end{equation}
Here $< \! R^{2}_{\! A} \! >$ is the radius squared,
$A^{}_{\rm eff \!, \; coh}$ takes account of nuclear screening
for the coherent process, and $i = L\/$ or $T\/$ designates the
polarization.
An experimentally based parameterization
of the
cross section was used for the 
production on the free nucleon, $\mu \, N \rightarrow \mu \, N \rho^{0}$.
Simulations were done as well \cite {asct} for incoherent $\rho^0$
production. 
  
 Coherent cross sections were generated for two models. For
the
complete color transparency
model (CT model) 
$A^{}_{\rm eff \!, \; coh}  = A\/$ was used. 
In another model, a substantial nuclear absorption (NA model) was assumed,
with
$A^{}_{\rm eff \!, \; coh} = A^{0.75 \!}$. 
The experiment measures the $t$-integrated coherent cross section, 
for which the CT model via integration of Eq. 1 predicts an approximate
$A^{4/3}$ dependence.
For the NA model, the
expected
$A$-dependence is weaker; $A^{5/6}$ similarly as for the
pion-nucleus
cross section.
For production of mesons not having the normal 
$q\bar{q}$ structure, 
the A-dependence of the cross section may be different. 
In that case, the A-dependence might even become a tool 
for investigating 4-quark or other exotic meson structure.

\section{RESULTS ON EXCLUSIVE $\rho^0$ \\PRODUCTION}
\label{lab_sec_4}

We considered 190 GeV muon beam. Simulations were 
done for carbon $(A = 12)$ and lead targets $(A = 207)$.
For each target we assumed the CT and the NA 
models.
The kinematic range considered for Q$^2$ and $\nu$, 
the energy of the virtual photon in the laboratory system,  
was:
\begin{equation} 2 < Q^2 < 20 \:\rm{GeV}^2,~~35 < \nu < 170 ~ \:\rm{GeV}. 
\label{EQ_MC6}
\end{equation}
We observe clear coherent peaks at small $t \approx p^{2}_{t}\/$
($< 0.05 \:\rm{GeV}^2$).
The numbers of accepted events for
the carbon and lead targets, assuming the two models
for the nuclear absorption, are given in Table \ref{table:1},
for a projected 150 day run.
 
\begin{table}[htb]
\caption{Number of accepted events for
the two models.}
\label{table:1}
\newcommand{\m}{\hphantom{$-$}}
\newcommand{\cc}[1]{\multicolumn{1}{c}{#1}}
\begin{tabular}{|c|c|c|}
\hline
  \raisebox{0mm}[4mm][3mm]{\hspace*{3mm} {\large model} \hspace*{3mm}}  &  {\large
\hspace*{6mm} $N^{}_{\rm C}$ \hspace*{6mm}}
   &  {\large \hspace*{6mm} $N^{}_{\rm Pb}$ \hspace*{6mm}}  \\
\hline
\hline
  \raisebox{0mm}[4mm][3mm]{\large CT}   &  {\large 70 000}  &  {\large 200 000}  \\
\hline
  \raisebox{0mm}[4mm][3mm]{\large NA}   &  {\large 28 000}  &  {\large 20 000}   \\
\hline
\end{tabular}\\[2pt]
\end{table}

As $Q^2$ increases, the approach to the CT limit is expected to be different
if the $\rho^0$ is produced by longitudinally or transversely polarized 
virtual photons.
We plan to 
study the $A$-dependence
of the cross sections for samples with different $\rho ^0$ polarizations,
which will be
selected by cuts on 
the 
$\cos \theta \/$ distributions for pions from $\rho^{0}$ decays.
These studies with {\it different polarizations\/} of the 
virtual photons are {\it important\/} for the clear demonstration of CT.

Another important aspect for CT studies is the covered
range
of the coherence length $l^{}_{c}\/$. 
It is important 
to disentangle effects due to CT from those caused by 
the modified absorption at small $l_c$ values. 
We may use the combined data at $l_c$ values exceeding the sizes
of the target nuclei, with 
the selection
$l^{}_{c} > l^{\rm min}_{c} \simeq 2 \cdot < r^{2}_{\rm Pb} >^{1/2} = 11 
\:\rm{fm}$.
About 50\% of events survive that cut on $l^{}_{c}$. They 
cover the range of $Q^{2} < 6~ \:\rm{GeV}^2 \!$, \,which is expected to be
sufficient to observe CT. 

We estimated values and 
statistical precision of $R^{}_{\rm T}$, the ratio of the nuclear transparencies
for lead and carbon, for different $Q^{2}$ bins
for $p^{2}_{t} < 0.02 \:\rm{GeV}^{2}$.
One expects large differences in $R^{}_{\rm T}$ for the two considered
models.
For coherent samples $R^{}_{\rm T} \approx 5$ for CT model and 
$\approx 1$ for NA model.
At $Q^2 \simeq 5 ~\:\rm{GeV}^2$ the precision of the measurement of
$R^{}_{\rm T}$ for coherent events will be better than 17\%, 
thus allowing excellent discrimination
between the 
models.

\section{COMPARISON WITH PREVIOUS EXPERIMENTS}
\label{lab_sec_5}

The 
NMC \cite {NMCJpsi},
and and E691 
\cite{Sokolov} 
cross sections 
for coherent J$/\Psi$ 
production have A-dependences close to 
A$^{4/3}$.  
The NMC and E665 data \cite {NMC,E665}
on incoherent $\rho^0$ production cover a large range
of $Q^2$ and  $l_c$ values.
Due to the moderate statistics of the data, 
the $Q^2$
dependence of nuclear 
absorption could not be obtained at sufficiently large or at fixed 
$l_c$, which may 
obscure the expected CT effects in incoherent production \cite{RPW,HERMES}. 
The NMC and E665  experiments published only limited  data \cite
{NMC,E665} for coherent
$\rho^0$ production. 

The COMPASS $\rho^0$ production data for C and Pb will cover
the kinematic range similar to that of the NMC $\rho^0$ production data,
and with
the high statistics given in Table 1. 
Due to large statistics, splitting of COMPASS data in several $Q^2$
and $l_c$ bins as well as the selection of coherent and incoherent events,
and  with longitudinal
or transverse $\rho ^0$ polarization, will be possible. 

\section{CONCLUSIONS}
\label{lab_sec_7}

We simulated
exclusive coherent $\rho^{0}$
muoproduction
$(\mu A \rightarrow \mu \, \rho^{0} A)$ events for the COMPASS experiment,
using thin nuclear targets of carbon and lead.
Good resolutions in $Q^{2} \!$, \,$l^{}_{c}$, $t$ ($p^{2}_{t}$) and $\cos
\theta \/$ are feasible.
An efficient selection of coherent events is possible
by applying cuts on $p^{2}_{t}$. In order to obtain the samples of events
initiated by either $\gamma^{\ast}_{L}\/$
or $\gamma^{\ast}_{T}\/$, the cuts on the $\rho^{0}$ decay
angular distribution of $\cos \theta \/$ will be used.
The search for CT
could be facilitated by using the events with $l^{}_{c}\/$ values exceeding
the sizes of the target nuclei. The fraction of such events is
substantial and the covered $Q^{2}$ range seems sufficient to observe CT.

We showed high sensitivity of the measured ratio $R^{}_{\rm T}$ 
for different models of nuclear absorption.
Good statistical accuracy 
may be achieved already during a 150 day run. 
These measurements, taken at different $Q^{2}$ intervals, may allow to
discriminate between different mechanisms of the interaction of the hadronic
components of the virtual photon with the nucleus, and 
should unambiguously demonstrate CT.


\end{document}